\documentstyle[preprint,aps]{revtex}

\begin{document}
\title{Local enhancement of spin and orbital antiferromagnetism around a single
vortex in high-Tc superconductors}
\author{J.An$^1$ and Chang-De Gong$^{2,1,3}$}
\address{$^1$National key laboratory of solid state microstructures, Department of
Physics,\\
Nanjing University, Nanjing 210093, China\\
$^2$Chinese Center of Advanced Science and Technology (World Laboratory),\\
P.O. Box 8730, Beijing 100080, China\\
$^3$Department of Physics, Chinese University of Hong Kong, Hong Kong, China.}
\date{\today }
\maketitle

\begin{abstract}
In a simple model we investigate three competing orders: d-wave
superconductivity, spin and orbital antiferromagnetism (AF) in the vicinity
of a single vortex in a cuprate superconductor. We find that when the
potential for the orbital AF $V_d$ is comparatively small, the
spin-density-wave (SDW) order has an enhancement at the vortex core center,
and the ''d-density-wave'' (DDW) order exhibits a rather weak behavior
around the core and vanishing small away from the core. However, when $V_d$
becomes large, globally, the SDW order decreases and the DDW order
increases; locally, not only the peak of the SDW order around the core still
exists, though relatively suppressed, but also a local peak for the DDW
order finally appears. Similar effects are also revealed for the features
when varying doping. Comparisons with experiments are discussed.
\end{abstract}


The microscopic structure around the field-induced vortices of high-Tc
cuprate superconductors (HTCS) is a challenging problem in condensed matter
physics. Physically, when an external magnetic field is applied to a HTCS,
an Abrikosov vortex lattice will be produced. By destroying the
superconducting (SC)\ order locally, in the small regions centered around
the vortex cores, the normal state is formed. Experimentally, muon spin
rotation measurements in underdoped YBa$_2$Cu$_3$O$_{6+x}$\cite{Miller},
nuclear magnetic resonance in optimally doped Tl$_2$Ba$_2$CuO$_{6+\delta }$ 
\cite{Kakuyanagi}have found the evidence for the presence of local AF in the
vortex cores. Furthermore, the presence of both local AF and charge order in
the vicinity of the field-induced vortices has been revealed recently by
neutron scattering experiments in optimally doped and underdoped La$_{2-x}$Sr%
$_x$CuO$_4$\cite{Lake}, scanning tunneling spectroscopy (STS) in Bi$_2$Sr$_2$%
CaCu$_2$O$_{8+\delta }$\cite{Hoffman}. Theoretically, the SO(5) theory\cite
{ZhangSC} has predicted an AF\ core in the underdoped cuprate
superconductors, where a superspin, combining the AF and the d-wave
superconductivity, rotate from lying inside the SC plane into AF sphere,
when the core is approached. On the other hand, staggered-current
correlations were predicted recently in lightly doped $t-J$ model\cite
{Ivanov,PWLeung}. In addition, based on the SU(2) gauge theory of the $t-J$
model, a staggered-current state for the vortex core has been proposed\cite
{PALee}, where the transition from the SC state to the staggered-current
state when approaching the core was interpreted in terms of the rotation of
an isospin vector, which has the meaning of the internal quantization axis
of the boson condensate. Other theories also predict in the vortex core
region the staggered current correlations\cite{WangQH}, charge order\cite
{Hoffman2,WangQH2}, both SDW and charge order\cite
{Franz,JXZhu1,ChenHD,ZhangY}, while a prediction of simultaneous presence of
charge, spin and orbital AF in the core region is still absent. The purpose
of this letter is to investigate this possibility in a unified framework. We
find that both the spin and orbital AF correlations has an enhancement at
the vortex core center. \qquad 

Here we assume the DDW state (or the staggered-current state, or the orbital
AF state) \cite{Chakravarty,Nayak} as the candidate state for the pseudogap,
which characterizes the normal state around the vortex cores. In addition,
it is well known that the competition between the d-wave superconductivity
and antiferromagnetism plays a rather prominent role in the mechanism of
superconductivity. Therefore, in the vortex core regions, at least three
ordering tendencies compete with one another and so are responsible for the
mechanism of the microscopic structure of the vortices. These orderings are
d-wave paring, spin AF, and the DDW order. We model our system as the
following effective mean-field Hamiltonian, which is an extension of the
model in reference\cite{JXZhu} in an external magnetic field,

\begin{eqnarray}
H &=&\{\sum_{<i,j>\sigma }(-t+(-1)^iW_{ij})-t^{^{\prime
}}\sum_{<i,j>^{^{\prime }}\sigma }\}\exp (i\varphi _{ij})c_{i\sigma
}^{+}c_{j\sigma }  \nonumber \\
&&\sum_i(U_{i\sigma }-\mu )c_{i\sigma }^{+}c_{i\sigma }+\sum_{<i,j>}(\Delta
_{ij}c_{i\uparrow }^{+}c_{j\downarrow }^{+}+H.c.)
\end{eqnarray}
\smallskip where $t$, $t^{^{\prime }}$ are the hopping integrals for the
nearest neighbors (n.n) $<i,j>$ and the next-nearest neighbors (n.n.n.) $%
<i,j>^{^{\prime }}$, respectively. $\varphi _{ij}=\frac \pi {\Phi _0}\int_{%
{\bf r}_j}^{{\bf r}_i}d{\bf r\cdot A(r)}$, where $\Phi _0=hc/2e$ is the
superconducting flux quantum and the external magnetic field ${\bf H}=\nabla
\times {\bf A}$ is directed along $z$ axis. The DDW order parameter is
defined as $W_{ij}=(-1)^i\frac{V_d}2\sum_\sigma <\exp (i\varphi
_{ij})c_{i\sigma }^{+}c_{j\sigma }-\exp (-i\varphi _{ij})c_{j\sigma
}^{+}c_{i\sigma }>$ , which is proportional to the physical current flowing
along the bond $ij$ and so it is naturally gauge invariant and then keeps
the gauge invariance of the Hamiltonian. $V_d$ is the potential for the DDW
channel, which favors a current-carrying state. $U_{i\sigma }=Un_{i-\sigma }$
is the spin-dependent scattering potential, where $U$ is the on-site
repulsion, favoring an antiferromagnetic SDW order.  $\Delta _{ij}=\frac{V_p}%
2<c_{i\uparrow }c_{j\downarrow }-c_{i\downarrow }c_{j\uparrow }>$ is the SC
spin-singlet pairing order parameter, where $V_p$ is the attractive pairing
interaction. $\mu $ is the chemical potential. Note that the above model can
be taken as the mean-field version of a more complicated and correlated
model, such as $t-U-V$ model. By decomposing the electron annihilating
operator into quasiparticles' creating and annihilating operators: $%
c_{i\sigma }=\sum_n(u_{n\sigma }(i)\gamma _{n\sigma }+v_{n\sigma }\gamma
_{n-\sigma }^{+})$, the above Hamiltonian can be diagonalized by solving the
following Bogoliubov-de Gennes (BdG) equations in a self-consistent way\cite
{WangY}, 
\begin{equation}
\left( 
\begin{array}{cc}
H_{ij}^{\uparrow } & \Delta _{ij} \\ 
\Delta _{ij}^{*} & -H_{ij}^{\downarrow *}
\end{array}
\right) \Psi (j)=E\Psi (i)
\end{equation}
where the quasiparticle wave function $\Psi (i)=(u_{\uparrow
}(i),v_{\downarrow }(i))^T$. The single-particle Hamiltonian reads $%
H_{ij}^\sigma =(-t+(-1)^iW_{ij})\delta _{i+{\bf \delta }},_j-t^{^{\prime
}}\delta _{i+{\bf \delta }^{^{\prime }}},_j+(U_in_{i-\sigma }-\mu )\delta
_{ij}$ with the subscript ${\bf \delta }=(\pm 1,0)$ or $(0,\pm 1)$, ${\bf %
\delta }^{^{\prime }}=(\pm 1,\pm 1)$ or $(\pm 1,\mp 1).$ The self-consistent
DDW and SC order parameters read respectively: $W_{ij}=(-1)^i\times iV_d%
\mathop{\rm Im}%
\sum_{n\sigma }\exp (i\varphi _{ij})(u_{n\sigma }(i)u_{n\sigma
}^{*}(j)+v_{n\sigma }(i)v_{n\sigma }^{*}(j))tanh(\beta E_n/2)$, and $\Delta
_{ij}$\ $=\frac{V_p}2\sum_n\sigma \{u_{n\sigma }(i)v_{n-\sigma }^{*}(j)-$\ $%
v_{n\sigma }^{*}(i)u_{n-\sigma }(j)\}tanh(\beta E_n/2)$. All the sum over $n$
is restricted to only positive eigenvalues $E>0$ for an excitation energy
should be positive-definite.

For simplicity, here the length and energy are measured in the units of the
lattice constant $a$ and the nearest-neighbor hopping $t$, respectively. To
extract the low-energy physics, we set the temperature to be zero. By using
the periodic boundary condition, the BdG equations are solved
self-consistently for a square lattice of $40\times 40$ sites. We have
chosen a symmetric gauge, where the vector potential can be written as ${\bf %
A=}\frac 12{\bf H}\times {\bf r}$ with ${\bf r}$ the radius vector measured
from the square center. The gauge phase $\varphi _{ij}$ on the link between
sites at two different edges of the boundary can be so chosen that all the
boundary effect be removed except at the four corners of the square. In the
region we are interested in, the corners' boundary has a rather small effect
on it and the results we have obtained should still be very convincing. The
starting state is the uniform d-wave superconducting state with vanishing
small staggered magnetization and staggered current. The d-wave symmetries $%
\Delta _{i,i+\delta }=\pm \Delta _0$, $W_{i,i+{\bf \delta }}=\pm W_0$, with $%
+$ for ${\bf \delta }=(\pm 1,0)$ and $-$ for ${\bf \delta }=(0,\pm 1)$, are
preserved at the beginning of the iterating.

In the whole article, we take the on-site energy $U=3$, the pairing
interaction $V_p=1.3$, the n.n.n hopping integral $t^{^{\prime }}=-0.25$.
The magnitude of ${\bf H}$ is so chosen that the total flux passing through
the square lattice is a flux quantum $\Phi _0$. The numerical results shows
that when the value of $V_d$ is relatively small, the DDW order will be
absent or be uniformly vanishing small. For $V_d=0.5$, we plot in Fig .1 the
spacial profiles of the vortex structure for the electron occupation number $%
n=0.86$. The site-dependent DDW order parameter $W_i$ is defined as $W_i=%
\frac 14(W$ $_{i,i+{\bf e}_x}+W_{i,i-{\bf e}_x}-W$ $_{i,i+{\bf e}_y}-W_{i,i-%
{\bf e}_y})$. Here $V_d$ is so weak that it can hardly stabilize a staggered
current. Actually, the staggered current is found to be nearly uniformly
distributed and vanishing small. The d-wave SC order parameter vanishes at
the vortex core center and tends to be finite when away from the core. The
gauge-invariant site-dependent SC order parameter $|\Delta _i|$ defined as $%
|\Delta _i|=\frac 14|\widetilde{\Delta }_{i,i+{\bf e}_x}+\widetilde{\Delta }%
_{i,i-{\bf e}_x}-\widetilde{\Delta }_{i,i+{\bf e}_y}-\widetilde{\Delta }%
_{i,i-{\bf e}_y}|$ with $\widetilde{\Delta }_{i,j}=\Delta _{ij}e^{i\varphi
_{ij}}$\cite{Discuss2}, weakly fluctuates and is not homogeneous even when
the distance from the core centre is larger than the superconducting
coherence length $\xi _0$. The antiferromagnetic SDW order parameter $%
m=(-1)^i(n_{i\uparrow }-n_{i\downarrow })$ (the staggered magnetization)
shows an enhanced peak at the vortex core center, exhibiting strong AF
correlation. The presence of another eight symmetric-positioned small peaks
around the vortex core implies the possibility of the quasi-periodic
magnetic structure with a much longer period $\xi _m$ (peak-peak separation)
than $\xi _0$. Here $\xi _m$ is a length scale of order about $8a\sim 10a$,
which can be compared to experiment\cite{Hoffman}. The electron
charge-density-wave (CDW) order parameter exhibits a similar modulation to
the SDW. The enhanced peak at the core center means the vortex carries a
negative charge. Another eight small peaks in the CDW appear at nearly the
same spatial places as the SDW. Note that in the trivial case of ${\bf H}=0$%
, the homogeneous SDW order increases by decreasing doping, or increasing
electron density. Surprisingly, this global characteristic seems to be
locally correct in this parameter region: the CDW form local peaks just
where the SDW have peaks.

In Fig. 2, we plot the same graphs for the same parameters except $V_d=0.65$%
. Now the strength of the DDW potential $V_d$ is so large that not only a
homogeneous small staggered current is stabilized, but also a great local
enhancement is formed at the core center. In another word, the
staggered-current correlations become much stronger when approaching the
vortex core from the outside. This local enhancement of the staggered
current can be interpreted as the effect of the local formation of the
normal pseudogap region through destroying the SC order around the vortex
core. This current-carrying state apparently breaks the time reversal
symmetry. This is consistent with the argument based on the $t-J$ model,
where an enhanced staggered current correlation when approaching the vortex
center was predicted\cite{PALee}. The d-wave SC order parameter approaches
zero at the vortex core center. Away from the vortex core, the SC order
parameter tends quickly to be a finite constant. Here the absence of the
weak modulation in the SC\ order parameter is associated with the
suppression of the modulation in the SDW order parameter. The SDW still
shows a strong peak (local maximum) at the core center, whereas the eight
small peaks surrounding the core are so suppressed that they are hardly
visible. This suppression of the magnetic modulation is due to the
competition between the SDW and DDW order parameters. Generally, the
presence of the DDW order frustrates the SDW order\cite{AnJ}, whereas the
presence of the SDW order seems to have a relative weak effect on the DDW
order. Therefore the DDW order leads to the suppression of the spin AF
correlations outside the vortex core. Inside the core, there exists two
competing effects: the DDW still suppresses the SDW, whereas the
disappearance of the SC order tends to increase the SDW. Combination of the
two competing effects causes the enhanced SDW order to be still present
inside the core, whereas the SDW order is strongly suppressed by the DDW\
order outside the core. The electron charge distribution (or CDW\ ) exhibits
a feature similar to that of the SDW, where a local maximum is formed at the
vortex core center, causing a negative vortex charge. The electrons are
accumulated locally by expelling out the holes at the vortex core, favoring
more strong spin AF\ correlations.

In order to make more clear the characteristic of the vortex structure when
the staggered-current correlations are dominant, we show the case $V_d=1$ in
Fig. 3. The strength of the DDW order is so strong that a nearly homogeneous
DDW order is induced, except that a relatively small local peak appears at
the core center. The magnitude of the DDW order is greatly increased, which
then cause a great suppression on the SDW order. The SDW order is nearly
homogeneous and vanishing small. A relatively small peak is still present
for the SDW order at the vortex center due to its characteristic as the
local normal state. The electron density is also nearly homogeneous except a
small broad local peak at the core center. The inset of Fig. 3 shows the
variation of the spatial averaged SDW (solid squares) and DDW (solid
circles) order parameters with $V_d$ for the same other parameters. $V_d$
has a threshold $V_d^c\approx 0.55$ , above which the DDW order appears and
begin to increase, while the SDW order begin to decrease.

Similar effects were found for the case where the doping level changes. We
show in Fig. 4 the spatial vortex structure with the electron occupation
number $n=0.91$ and $n=0.94$, respectively. Combined with Fig.2, it can be
seen that small doping strongly favor a SDW\ order, both globally, and
locally at the vortex core. The local peak of the SDW\ order is strongly
enhanced. This great local enhancement comes from not only the doping
effect, but also the normal state characteristic by destroying the SC order
locally at the vortex core. The two effects have the same tendencies to
increasing the SDW order, so the relatively great local enhancement is well
explained. A quasi-periodic magnetic structure also appears in the SDW,
indicating that this is not the feature just belonging to the optimal
system, which is consistent with the observation by the neutron scattering
experiments\cite{Lake}. The relative local DDW peak at the core center
decreases when decreasing doping. This can be explained if we note that (i)
zero doping stabilizes a strong SDW order and a much smaller DDW order
because of the on-site energy, which depresses the ability of the electrons'
hopping between sites, and then the DDW order can be seen to have a tendency
to decrease when decreasing doping; (ii) the vortex core is a hole-poor
region, relative to the region outside the core. Due to the same reason that
the core has the nature of the normal state, the local peak of the\ DDW
survives, though rather small, when decreasing doping. The CDW\ order always
has an enhanced peak at the core center when doping level changes, and as a
consequence, it is always indicating a negative vortex charge in the
parameter region considered, in contrast to that of Chen\cite{ChenY}, where
a change of sign of the vortex may be possible, when vortex cores without AF
are formed. This may be due to the large Coulomb on-site energy we have
chosen, so that the core without AF is never stabilized.\ \ \ \ \ \ \ \ \ \
\ \ \ \ \ \ \ 

This work is supported by the Chinese National Natural Science Foundation.

\begin{figure}[tbp]
\caption{The spatial distribution of (a) the d-wave SC order parameter, (b)
electron density, (c) the staggered magnetization, (d) the DDW order
parameter for a $40\times 40$ lattice. Only the center $34\times 34$ lattice
is shown to eliminate the corners' boundary effect. Parameter values: $%
t^{^{\prime }}=-0.25$, $U=3$, $V_p=1.3$, $V_d=0.5$ and $n=0.86$.}
\end{figure}
\begin{figure}[tbp]
\caption{The same as Fig. 1 except $V_d=0.65$.}
\end{figure}
\begin{figure}[tbp]
\caption{The same as Fig. 1 except $V_d=1.0$.}
\end{figure}
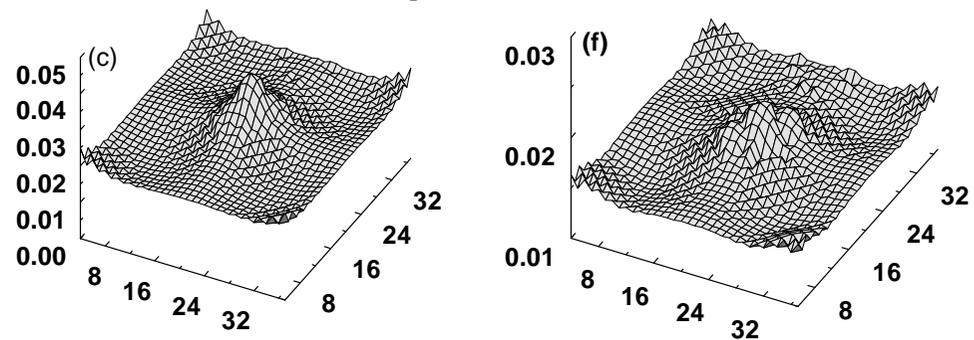
\begin{figure}[tbp]
\caption{The spatial distribution of (a), (c) electron density; (b), (d) the
staggered magnetization; (c), (f) the DDW order parameter. Parameters have
the same values as Fig. 2 except n=0.91 for (a), (b), (c), whereas n=0.94
for (d), (e), (f).}
\end{figure}


\begin{references}
\bibitem{Miller}  R. I. Miller {\it et al., }Phys. Rev. Letts {\bf 88},
137002 (2002).

\bibitem{Kakuyanagi}  K. Kakuyanagi, K. Kumagai, Y. Matsuda, and M.
Hasegawa, Phys. Rev. Letts {\bf 90}, 197003 (2003).

\bibitem{Lake}  B. Lake {\it et al., }Science {\bf 291, }1759 (2001); B.
Lake {\it et al.,} Nature (London) {\bf 415}, 299 (2002).

\bibitem{Hoffman}  J. E. Hoffman {\it et al.}, Science {\bf 295},466(2002).

\bibitem{ZhangSC}  S. C. Zhang, Science {\bf 275}, 1089 (1997); D. P.
Arovas, A. J. Berlinsky, C. Kallin, S. C. Zhang, Phys. Rev. Lett {\bf 79},
2871 (1997).

\bibitem{Ivanov}  D. A. Ivanov, P. A. Lee, and X. G. Wen, Phys. Rev. Letts 
{\bf 84,} 3958(2000).

\bibitem{PWLeung}  P. W. Leung, Phys. Rev. B {\bf 62}, 6112(2000).

\bibitem{PALee}  P. A. Lee, X. G. Wen, Phys. Rev. B {\bf 63}, 224517 (2001);
J. Kishine, P.\ A. Lee and X. G. Wen, Phys. Rev. Letts, {\bf 86}, 5365
(2001).

\bibitem{WangQH}  Q. H. Wang, J. H. Han, D. H. Lee, Phys. Rev. Letts {\bf 87}%
, 167004 (2001).

\bibitem{Hoffman2}  J. E. Hoffman, K. McElroy, D.-H. Lee, K. M. Lang, H.
Eisaki, S. Uchida, J. C. Davis, Science {\bf 297}, 1148 (2002).

\bibitem{WangQH2}  Q.\ H. Wang, D. H. Lee, Phys. Rev. B {\bf 67}%
,020511(2003).

\bibitem{Franz}  M. Franz, D. E. Sheehy, and Z. Te\v{s}anovi\'{c}, Phys.
Rev. Letts {\bf 88}, 257005 (2002); M. Franz, Z. Te\v{s}anovi\'{c}, Phys.
Rev. B {\bf 63}, 064516 (2001).

\bibitem{JXZhu1}  J. X. Zhu, I. Martin, A. R. Bishop, Phys. Rev. Letts {\bf %
89}, 067003 (2002); J. X. Zhu, C. S. Ting, Phys. Rev. Letts {\bf 87}, 147002
(2001).

\bibitem{ChenHD}  H. D. Chen, J. P. Hu, S. Capponi, E. Arrigoni, S. C.
Zhang, Phys. Rev. Letts {\bf 89}, 137004 (2002).

\bibitem{ZhangY}  Y. Zhang, E. Demler, S. Sachdev, Phys. Rev. B {\bf 66},
094501 (2002).

\bibitem{Chakravarty}  S. Chakravarty, R. B. Laughlin, D. K. Morr, C. Nayak,
Phys. Rev. B {\bf 63}, 094503 (2001). 

\bibitem{Nayak}  C. Nayak, Phys. Rev. B {\bf 62}, 4880 (2000); C. Nayak,
Phys. Rev. B {\bf 62}, 6135 (2000).

\bibitem{JXZhu}  J. X. Zhu, W. Kim, C. S. Ting, J. P. Carbotte, Phys. Rev.
Letts {\bf 87}, 197001 (2001).

\bibitem{WangY}  See for example, Y. Wang and A. H. MacDonald, Phys.\ Rev. B 
{\bf 52}, 3876 (1995); P. I. Soininen, C. Kallin, A. J. Berlinsky, Phys.
Rev. B {\bf 50}, 13883 (1994).

\bibitem{Discuss2}  In an external magnetic field, the conventional
definition $|\Delta _i|=\frac 14|\Delta _{i,i+{\bf e}_x}+\Delta _{i,i-{\bf e}%
_x}-\Delta _{i,i+{\bf e}_y}-\Delta _{i,i-{\bf e}_y}|$ is not invariant in
the gauge transformations $c_{i\sigma }\rightarrow c_{i\sigma }e^{-i\theta
_i}$, $\Delta _{ij}\rightarrow \Delta _{ij}e^{-i(\theta _i+\theta _j)}$, $%
\varphi _{ij}\rightarrow \varphi _{ij}-\theta _i+\theta _j$.

\bibitem{AnJ}  J. An and C. D. Gong, Phys. Rev. B {\bf 63}, 174434 (2001).

\bibitem{ChenY}  Y. Chen, Z. D. Wang, J. X. Zhu, C. S. Ting, Phys.\ Rev.
Letts {\bf 89}, 217001 (2002).
\end{references}
\end{document}